\documentclass{appolb}
\usepackage{color}
\usepackage{amsmath}
\usepackage{amssymb}
\usepackage[dvips]{graphicx}
\usepackage{subfig}
\usepackage{graphicx}

\newcommand{\be}{\begin{equation}}
\newcommand{\ee}{\end{equation}}

\newcommand{\uvec}[1]{\underline{#1}}

\begin{document}
\title{Gluon Wavefunctions and Amplitudes on the Light-Front%
\thanks{Presented at the "Light Cone 2012" conference, Krak\'ow, Poland, July 8-13, 2012.}%
}
\author{Christian A. Cruz-Santiago
\address{Physics Department, Penn State University, 104 Davey Laboratory\\ University Park, PA 16802, USA}
\\
}
\maketitle
\begin{abstract}
We investigate the  tree level multi-gluon components of the gluon light cone wavefunctions in the light cone gauge keeping the exact kinematics of the gluon emissions. We focus on the components with all helicities identical to the helicity of the incoming gluon.
The recurrence relations for the gluon wavefunctions are derived.
In the case when the virtuality of the incoming gluon is neglected the exact form of the multi-gluon wavefunction as well as the fragmentation function is obtained.
Furthermore we analyze the 2 to N tree-level gluon scattering in the framework of light-front perturbation theory and we demonstrate that the amplitude for this process can be obtained from the 1 to N+1 gluon wavefunction.    Finally, we demonstrate that our results for selected helicity configurations are equivalent to the Parke-Taylor amplitudes.
\end{abstract}
  
\section{Introduction}
\begin{figure}[h]
\centering
\subfloat[]{\includegraphics[width=.4\textwidth]{{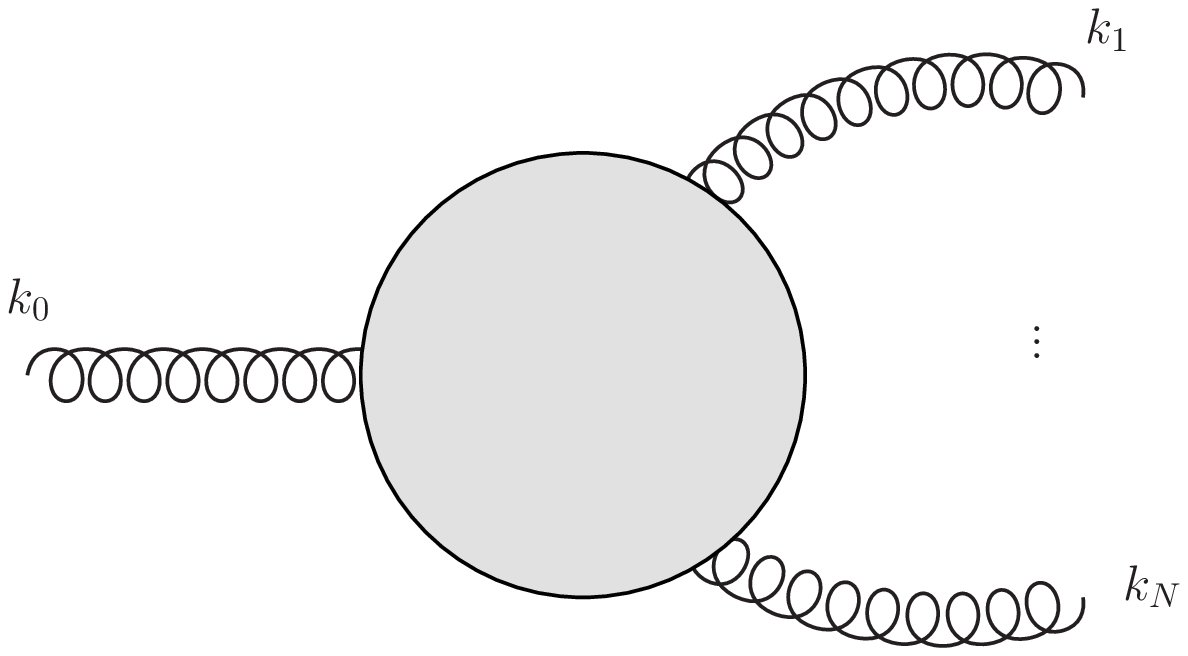}}}\hspace*{2cm}
\subfloat[]{\includegraphics[width=.4\textwidth]{{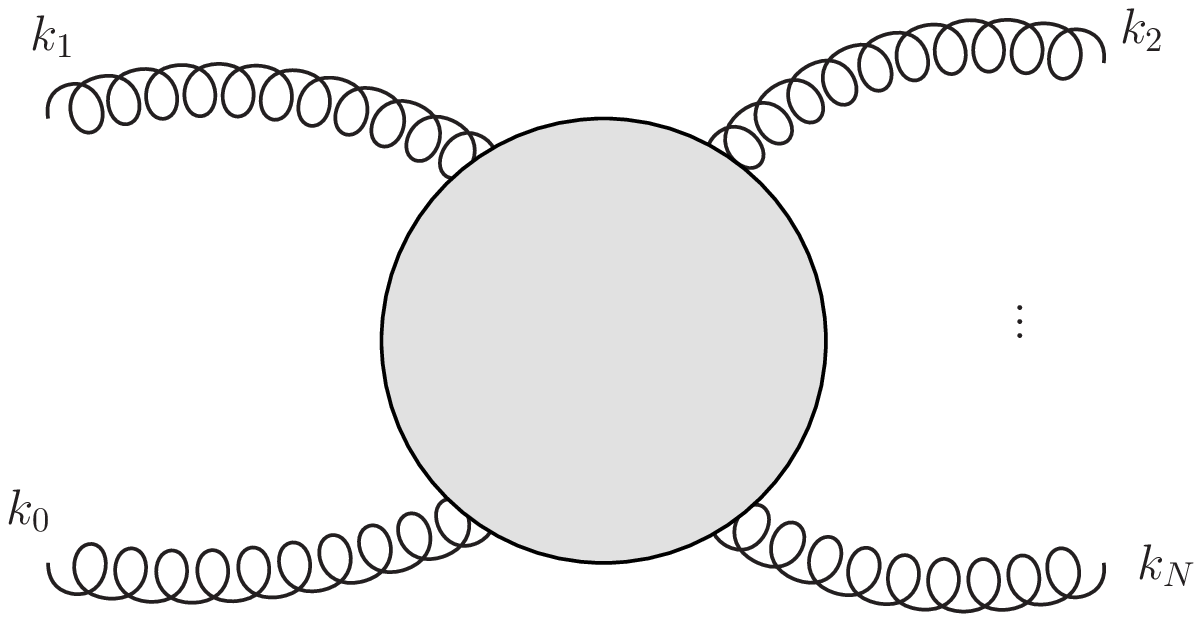}}}
\caption{The two types of tree-level gluon diagrams that are explored in this paper.  Light-cone time is forward to the right.}
\label{fig:1&2toN}
\end{figure}

Light-front quantization \cite{Dirac:1949cp} was first proposed by Dirac in 1949 as a new way of doing quantum field theory.  The purpose of this paper is to use this tool (specifically, the light-front perturbation theory) to explore gluon wavefunctions, fragmentation functions and scattering amplitudes, see for example Fig.~\ref{fig:1&2toN}.  In particular, we aim to determine the recursion relations between these objects and some of their properties. Here we will be following the work done by Motyka and Sta\'sto in \cite{ms} and expanding upon it.

Before continuing, there are two things which we need to remark.  Throughout this paper we will be using the Lepage-Brodsky convention \cite{Brodsky:1997de}.  If a 4-vector $a^\mu$ is defined in instant form by $a^\mu = (a^0,a^1,a^2,a^3))$ then in the light-front it is defined as $a^\mu=(a^+,a^1,a^2,a^-)$, where $a^+= a^0+a^3$ and $a^-=a^0-a^3$.  Also, when referring to amplitudes we are actually talking about color-ordered helicity sub-amplitudes \cite{Mangano:1990by}.  The Parke-Taylor amplitudes \cite{Mangano:1990by,Parke:1986gb} are given by
${\cal M}_n \;=\; \sum_{\{1,\dots,n\}} {\rm tr}(t^{l_1}t^{l_2}\dots t^{l_n}) \; m(k_1,\lambda_1;k_2,\lambda_2;\ldots;k_n,\lambda_n) \; $,
where $l_1,l_2,\ldots,l_n$, $k_1,k_2,\ldots,k_n$ and $\lambda_1,\lambda_2,\ldots,\lambda_n$ are the color indices, momenta and the helicities of the external $n$ gluons, respectively. Matrices $t^a$ are in the fundamental representation of the color $SU(N_c)$ group. The sum is over the $(n-1)!$ non-cyclic permutations of the set $\{0,1,\dots,n\}$.  Finally, $m$ are the color-ordered helicity sub-amplitudes that we will be referring to.  These amplitudes obey a variety of properties out of which the most relevant to us is that they depend only on the kinematics of the system and are color-independent.

\begin{figure}[h]
\centering
\subfloat[]{\includegraphics[width=.3\textwidth]{{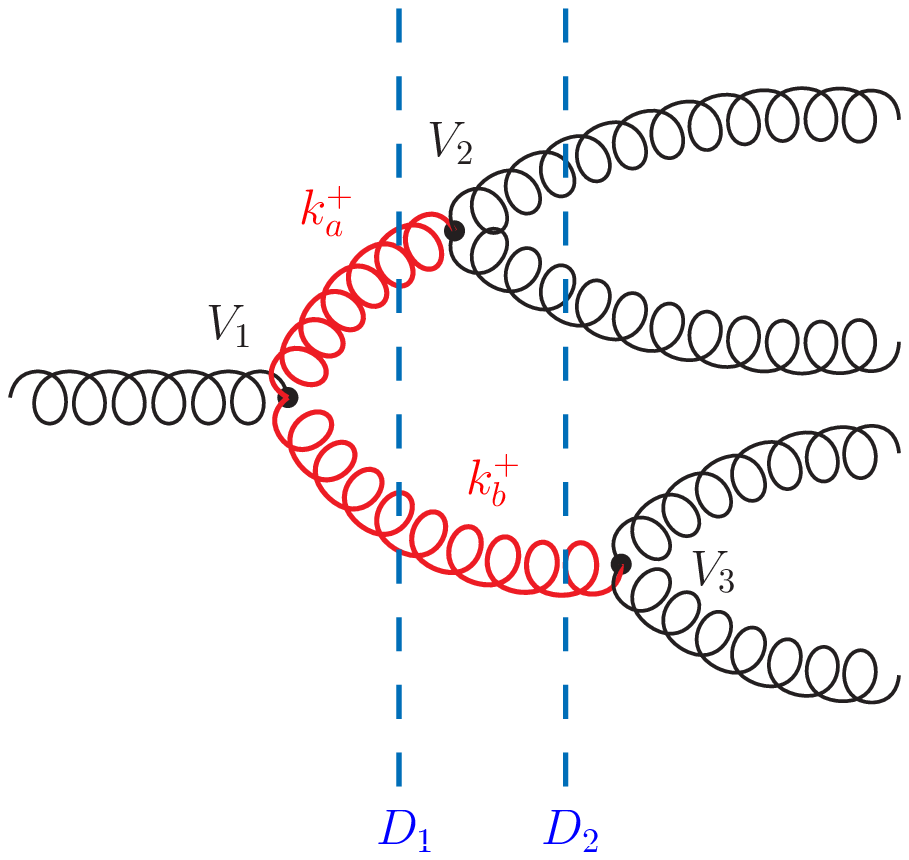}}\label{fig:vertexOrd1}}\hspace*{2cm}
\subfloat[]{\includegraphics[width=.3\textwidth]{{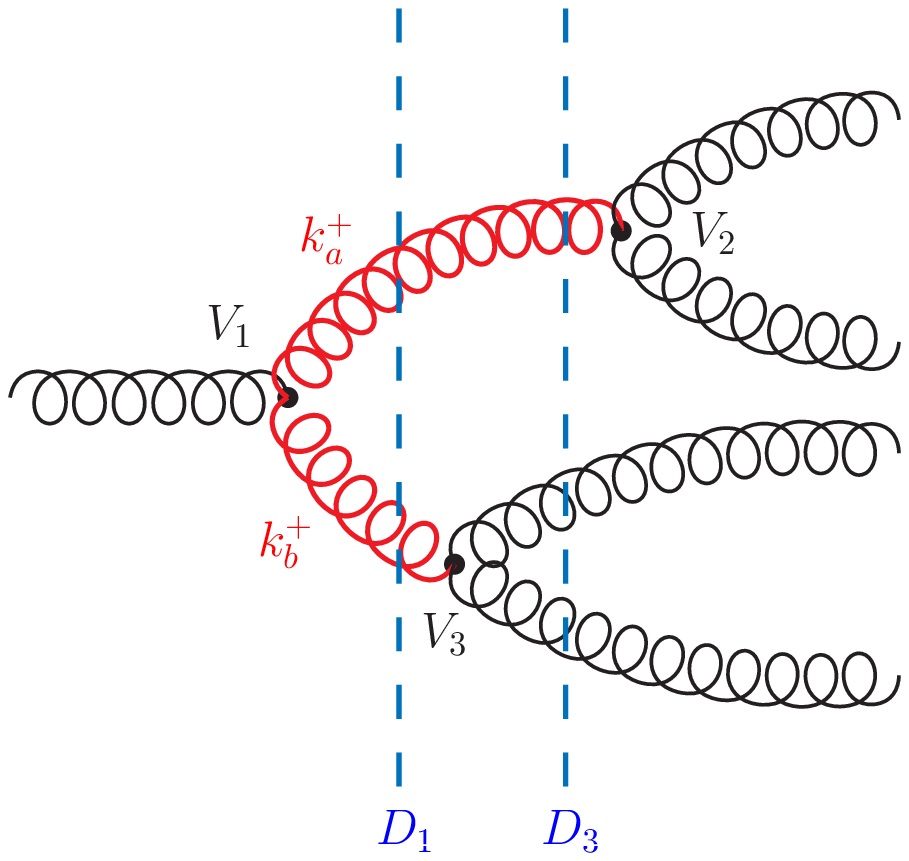}}\label{fig:vertexOrd2}}
\caption{Graphs contributing to the 1 to 4 transition.  Shown are the two different vertex orderings.}
\label{fig:vertexOrd}
\end{figure}

To calculate wavefunctions, fragmentation functions and amplitudes we need to know the light-front rules \cite{Brodsky:1997de,Kogut:1969xa}.  
First, for each vertex, internal line and intermediate state we get a factor $V$, 	$k^+$ and $D$ respectively.  So, for example, for Fig.~\ref{fig:vertexOrd1} the amplitude $M_a$ would be given by $M_a= {\prod V \over {\color{red} \prod k^+ }{\color{blue}\prod D}}$. Here $D$ is the energy denominator and it is given by $D= \sum_i k_i^- - \sum_j k_j^-$, whereas $V$ will depend on the choice of helicities.  The sum over $i$ is the over the initial state and $j$ is over the $n$th intermediate state.
	 Second, sum over all possible vertex orderings.  So, in our example we would not have only Fig.~\ref{fig:vertexOrd1}, we would also have Fig.~\ref{fig:vertexOrd2}.  Then, the total amplitude will be the sum of these two contributions. 
Third, all $k^+>0$.  So diagrams with lines going backwards in light-cone time will vanish.
	Finally, longitudinal and transverse momenta are conserved.  However, we should note that light cone energy is not conserved except for final and initial states.

\section{Gluon wavefunctions}
\begin{figure}[ht]
\centerline{\includegraphics[width=0.4\textwidth]{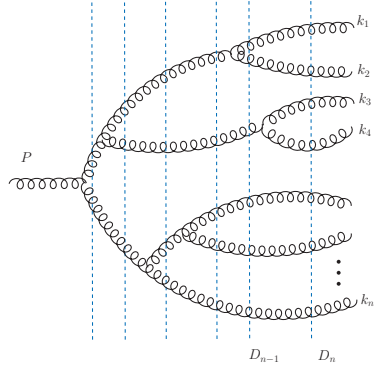}}
\caption{Graph contributing to the 1 to $N$ gluon wavefunction. }
\label{fig:onium_n}
\end{figure}

In this section and the next we will be closely following the work done in \cite{ms}.  To calculate the wavefunction $\Psi$ one applies the light-front rules to the object in Fig.~\ref{fig:onium_n} and all its relatives, and sum.  By relatives we mean the other tree-level graphs with the same initial and final state properties (number of gluons and helicity configuration).  This procedure is very similar to the one for calculating the transition from 1 to $N$.  However, there is a key difference:  in the wavefunctions the initial gluons are on-shell but the final gluons are off-shell.  Thus, the gluons in the final state can still interact.

For the case when the incoming and outgoing gluons have positive helicities one can find a recursion relation for the wavefunctions \cite{ms}.  The recursion relation is found to be, using $V= (z_i+z_{i+1})\uvec{\epsilon}^{(-)}\cdot \uvec{v}_{i\;i+1}$ because of the helicity choices,
\be
\Psi_{N+1}(k_1,k_2,\ldots,k_{N+1}) =\sum_{i=1}^N \frac{\uvec{\epsilon}^{(-)}\cdot \uvec{v}_{i\;i+1}}{\sqrt{\xi_{i\;i+1}}D_{N+1}} \Psi_{N}(k_1,\ldots,k_{i\;i+1},\ldots,k_{N+1}),
\label{eq:recurRel}
\ee
where $\uvec{\epsilon}^{(-)}$ is the tranverse polarization vector, $z$ is the longitudinal momentum fraction,
${\uvec{v}}_{(i_1 i_2 \ldots i_p)(j_1 j_2 \ldots j_q)} = 
{{\uvec{k}}_{i_1} + {\uvec{k}}_{i_2} + \ldots + {\uvec{k}}_{i_p}
\over z_{i_1} + z_{i_2} + \ldots + z_{i_p}} -
{{\uvec{k}}_{j_1} + {\uvec{k}}_{j_2} + \ldots + {\uvec{k}}_{j_q}
\over z_{j_1}+ z_{j_2} + \ldots + z_{j_q}}\; $ and
$\xi_{(i_1 i_2 \ldots i_p)(j_1 j_2 \ldots j_q)} = 
{
(z_{i_1}+z_{i_2}+\ldots+z_{i_p})
(z_{j_1}+z_{j_2}+\ldots+z_{j_q})
\over 
z_{i_1}+ z_{i_2}+ \ldots+ z_{i_p} + z_{j_1} + z_{j_2} + \ldots + z_{j_q}} \;.$
Recursion \eqref{eq:recurRel}  can be solved exactly to give:
\begin{align}
\Psi_{N+1}&(k_1,k_2,\ldots,k_{N+1}) = \frac{g^N}{\sqrt{z_1z_2\ldots z_{N+1}}}\frac{1}{\xi_{(12\ldots N)\;N+1}v_{(12\ldots N)\;N+1}} \nonumber \\
&\times\frac{1}{\xi_{(12\ldots N-1)(N\;N+1)}v_{(12\ldots N-1)(N\;N+1)}}\ldots\frac{1}{\xi_{1(2\ldots N+1)}v_{1(2\ldots N+1)}}
\label{eq:wavefunction}
\end{align}



\section{Gluon fragmentation function}
These are somewhat similar to the wavefunctions.  Whereas in the wavefunction the initial state is on-shell and the final off-shell, here the initial state is off-shell and the final on-shell.  Thus, these represent the final scatterings of gluons.  Just as in the case of the wavefunctions, one can also write a recursion relation which can be solved explicitly (for the same helicity configuration as before) for the fragmentation functions.  It is shown in \cite{ms} that these are given by
\be
T[(12\ldots N)\to 1,2,\ldots,N]=g^{N-1}\left(\frac{z_{(12\ldots N)}}{z_1z_2\ldots z_N}\right)^{3/2}\frac{1}{v_{12}v_{23}\ldots v_{N-1\;N}}
\label{eq:frag}
\ee


\section{2 to $N$ scattering amplitude}

\begin{figure}[h]
\centering
\includegraphics[width=.8\textwidth]{{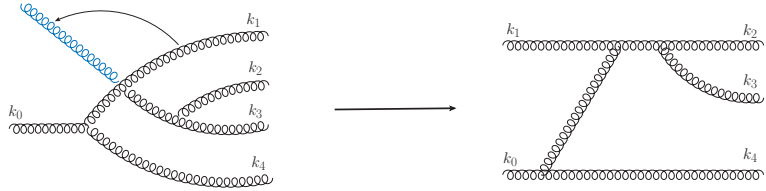}}
\caption{Shows the change of gluon 1 from outgoing to incoming}
\label{fig:conversion}
\end{figure}

The goal of this section is to show how the 2 to $N$ scattering amplitude can be derived from the wavefunctions.  The first step towards finding the amplitude is finding all the graphs that contribute to it.  To do this we can take each of the graphs that contribute towards the wavefunction and convert the $k_1$ gluon to be incoming, as in Fig.~\ref{fig:conversion}.  When doing this one must invert the helicity of gluon 1 and make sure to include all the other possible vertex orderings.  This, of course, means that the number of graphs contributing to the 2 to $N$ amplitude is much larger than for the wavefunction.  However, we can reduce the number of graphs.  If, for example, we have a group of graphs of the same topology, but some of them differ by the direction of some of the internal lines, we can make subgroups from graphs with the same line directions.  Then, the total contribution of the group will be equal to any of its subgroups \cite{tbp}.  Another interesting thing is that one of the subgroups is composed solely of graphs which come directly from the wavefunction (i.e. different vertex orderings are not included).  Thus, we can conclude that there is a one to one correspondence between the graphs that contribute to the 2 to $N$ amplitude and those which contribute to the wavefunction.

Now that we have all the graphs we can apply the light-front rules to each and sum.  However, if we know the wavefunction there is an easier way.  Let us compare a graph from the 2 to $N$ amplitude with its equivalent from the wavefunction.  There are three things to look at: vertex and internal line factors, and energy denominators.  For the vertex factor it is fairly easy to see that only the vertex with the switched gluon brings a change: an extra negative. E.g., whereas we have $v_{21}$ for the amplitude, we will have $-v_{21}$ for the wavefunction.  

For the internal line factor, let us look at Fig.~\ref{fig:vertexOrd}.  If starting from the left and going clockwise the external gluons are labelled 0, 1, 2, 3, 4,  then $z_a=z_1+z_2$.  However, if we flip the direction of gluon 1 then $z_a=-z_1+z_2$.  Nevertheless, we can make the following definition: $k_A \equiv k_1$ and $k_A \equiv -k_1$ for the wavefunction and the amplitude respectively.  With this we can write $z_a = z_A + z_2$ in both cases.  All internal lines can be written in terms of $z_A$ and it can be shown that the expressions are always the same for both cases.  

For the energy denominator factor let us look at $D_2$ in Fig.~\ref{fig:vertexOrd1}.  It can be written as $D_2={}^{\uvec{k}_0^2}/_{k_0^+}-{}^{\uvec{k}_1^2}/_{k_1^+}-{}^{\uvec{k}_2^2}/_{k_2^+}-{}^{\uvec{k}_b^2}/_{k_b^+}$.  But if we flip gluon 1 $D_2={}^{\uvec{k}_0^2}/_{k_0^+}+{}^{\uvec{k}_1^2}/_{k_1^+}-{}^{\uvec{k}_2^2}/_{k_2^+}-{}^{\uvec{k}_b^2}/_{k_b^+}$.  Just as in the case of the internal lines we can use $k_A$ to write $D_2={}^{\uvec{k}_0^2}/_{k_0^+}-{}^{\uvec{k}_A^2}/_{k_A^+}-{}^{\uvec{k}_2^2}/_{k_2^+}-{}^{\uvec{k}_b^2}/_{k_b^+}$.  Once again, when writing all denominators in terms of $k_A$ we arrive at the same expressions for both cases.  However, we must remember that the final state of the wavefunction is off-shell.  Overall this means that the  wavefunction's energy denominator factor will be the same as the amplitude's except for an extra factor of $1/(\sqrt{z_0z_1\ldots z_N}D_N)$, where $D_N$ is the final state's energy denominator.

Seeing as how there is a one to one correspondence between the graphs that contribute to the 2 to $N$ amplitude $M_{2\to N}$ and those that contribute to the 1 to $N+1$ wavefunction $\Psi_{N+1}$ we can conclude that the following relationship exists
\begin{align}
M_{2\to N}&(\{\uvec{k}_0,z_0;\uvec{k}_1,z_1\};\{\uvec{k}_2,z_2;\ldots;\uvec{k}_N,z_N\})  = \nonumber \\ 
-&\sqrt{z_0z_A\ldots z_N}D_N\Psi_{N+1}(k_0;k_A,k_2,\ldots,k_{N+1}) |_{\uvec{k}_A\to-\uvec{k}_1,z_a\to -z_1,D_N\to0}.
\label{eq:M&Psi}
\end{align}
This relationship is true for any helicity configurations, as we have made no assumptions about helicities when deriving it.  Nevertheless we can see how it works with the helicity configuration chosen in section 2.  For the amplitude this would mean the incoming and outgoing gluons would have helicities of $+-$ and $+\ldots+$ respectively.  For this configuration the wavefunction is given by \eqref{eq:wavefunction}.  Looking at it, we see that for this particular configuration the wavefunction cannot cancel the $D_N$ in \eqref{eq:M&Psi}.  Thus, when taking the limit $D_N\to 0$, the entire right-hand side of \eqref{eq:M&Psi} vanishes and, therefore, $M_{2\to N} = 0$ for the selected helicities.  This result matches the Parke-Taylor amplitude for the same helicity configuration.

\section{Summary and outlook}
In summary I have presented the wavefunctions, fragmentation functions and their recursion relations.  I have also presented how the wavefunctions can be used to find the 2 to $N$ amplitudes.  In future work we wish to confirm the MHV 2 to $N$ amplitudes and compute amplitudes for other, more complicated, helicity configurations. 


\end{document}